# An Agent-Based Intrusion Detection System for Local Area Networks

Jaydip Sen

Innovation Lab, Tata Consultancy Services Ltd.
Bengal Intelligent Park, Salt Lake Electronic Complex, Kolkata 700091, INDIA
Jaydip.Sen@tcs.com

*Abstract*: Since it is impossible to predict and identify all the vulnerabilities of a network beforehand, and penetration into a system by malicious intruders cannot always be prevented, *intrusion detection systems* (IDSs) are essential entities to ensure the security of a networked system. To be effective in carrying out their functions, the IDSs need to be accurate, adaptive, and extensible. Given these stringent requirements and the high level of vulnerabilities of the current days' networks, the design of an IDS has become a very challenging task. Although, an extensive research has been done on intrusion detection in a distributed environment, distributed IDSs suffer from a number of drawbacks e.g., high rates of false positives, low detection efficiency etc. In this paper, the design of a distributed IDS is proposed that consists of a group of autonomous and cooperating agents. In addition to its ability to detect attacks, the system is capable of identifying and isolating compromised nodes in the network thereby introducing fault-tolerance in its operations. The experiments conducted on the system have shown that it has a high detection efficiency and low false positives compared to some of the currently existing systems.

## 1. Introduction

There have been two different approaches for securing networks and host computers from malicious attackers: i) intrusion prevention mechanisms that include cryptographic techniques to safeguard sensitive information from unauthorized access, ii) intrusion detection mechanisms that recognize an ongoing attack on a system and respond appropriately to thwart such intrusive attempts. An *intrusion detection system* (IDS) is a security mechanism that can monitor and detect intrusions to the computer systems in real time. An IDS can be either host-based (sources of data are operating systems and applications audit trails), or network-based (monitor and analyze network traffic), or a combination of both these types. Conventional approaches to intrusion detection involving a central unit to monitor an entire system have several disadvantages [1]. To circumvent the demerits of a centralized IDS, the research in the field of intrusion detection over the last decade has been heading towards a distributed framework of monitors that do local detection and provide information to perform global detection of intrusions. In these systems, the local intrusion detection components look for local intrusions and pass the analysis of their results to the upper levels of the hierarchy. The components at the upper levels analyze the refined data from multiple lower level components and seek to establish a global view of the system state. Gopalakrishna and Spafford [1] argue that such an IDS is not really a distributed system because data analysis activity is performed at the nodes situated at the higher level of hierarchy in a centralized fashion. In such systems, transfer of data among the nodes can be a problem as it can lead to security breaches. Moreover, these systems suffer from the drawback of a single point of failure. An intruder can take control of the whole system if he or she can compromise the central server. Thus design of a distributed IDS is a challenging task and a number of issues are to be taken into consideration for this purpose, e.g.: reduction of the false positives, sufficient protection against compromised nodes, a secure communication mechanism between the distributed components etc.

In this paper, the scheme of a distributed IDS is presented. This is an extension of our earlier work presented in [2]. The IDS consists of a large number of autonomous agents that cooperate with each other for detecting any intrusive activity in the system. The system uses coordinated surveillance by incorporating inter-agent communication and distributed computing in decision making to identify early signs of an attack and recognize situations that are likely to happen before an attack takes place (e.g., systematic scanning of the network resources). It also raises an appropriate alarm whenever an attack is detected. The two primary goals of the proposed security mechanism are:

- *Detection of intrusive activities*: In addition to its ability to respond to an attack, the system should be able to alert the system administrator whenever it finds any sign of pre-attack activities. By local monitoring and sharing individual belief-estimates, the agents in the system can recognize and preempt activities that resemble security threats.
- *Identifying and isolating compromised hosts*: The system should be capable of detecting and isolating compromised nodes. This feature is incorporated in the system by implementing a distributed trust mechanism between the communicating agents.

Any IDS is required to handle uncertainties related to the domain and environment in which it works. To model this uncertainty, the agents in the proposed system represent their knowledge about attacks scenarios in the form of Bayesian networks [3]. This knowledge is introduced into the agents from analysis of repositories of data related to network attacks. The knowledge is distributed so that each agent is required to monitor only a relatively few aspects of the local network. However, the agents share their beliefs and through timely coordination of the agents, the system is able to detect more complex distributed attacks. To enable such distributed inference, the concept of *multiply sectioned Bayesian networks* (MSBN) [4] is utilized for representation of domain knowledge. The clique-tree propagation algorithm [5] is used for reasoning. To reduce network congestion and message overhead, the agents are grouped into sub-domains (localities), so that majority of the communications among the agents are within the sub-domains only.

The organization of the paper is as follows. Section 2 presents some existing work on distributed intrusion detection systems. Section 3 provides some background information about Bayesian networks and MSBNs. Section 4 describes the architecture of the proposed system and the agents constituting the IDS. Section 5 presents a brief overview of the communication mechanism and security of communication among the agents in the system. Section 6 describes how the concept of Bayesian networks is applied in intrusion detection and how inferences can be drawn using an MSBN framework. Section 7 presents a distributed trust management among the peer hosts that provides fault-tolerance in the proposed security mechanism. Section 8 gives the details of experiments conducted on the prototype architecture and the results obtained. Finally, Section 9 concludes the paper identifying some future scope of work.

## 2. Related Work

The approach of distributed intrusion detection is not new. Many researchers have proposed systems based on distributed detection of intrusions in large networks. In this section, some of these schemes are discussed briefly.

Snapp et al proposed DIDS - a distributed intrusion detection system that consists of host managers and LAN managers for distributed data monitoring and sending of notable events to the DIDS director [6]. The managers also do some local detection and pass the summaries to the director. At the local level, DIDS uses both statistical and rule-based detection, and at the global level it uses a rule-based expert system. The director analyzes the events to determine the security state of the system. This centralized nature of the director is clearly the bottleneck of the distributed approach of DIDS.

Porras and Neumann have presented a framework called EMERALD for distributed intrusion detection [7]. It employs monitors at the levels of hosts, domains and enterprises to develop an analysis hierarchy. It uses subscription-based communication scheme, both within and between the monitors. However, the inter-monitor subscription scheme is hierarchical in nature and limits access to the events or results from the layer immediately below.

AAFID is a distributed intrusion detection system developed in CERIAS at Purdue University [8][9]. AAFID employs agents at the lowest level of the hierarchy for data collection and analysis and transceivers and monitors at the higher levels for controlling agents and obtaining a global view of activities. It also uses filters within the hosts for identifying the relevant observations for intrusion detection. Essentially, the filters are data selection and abstraction layers, which provide a subscription-based service to the agents.

Dickerson et al have proposed a *fuzzy intrusion recognition engine* (FIRE) for network intrusion detection that uses fuzzy systems to assess malicious activity inside a computer networks [10]. It utilizes AAFID architecture as the base platform.

Qin et al have proposed deployment of a number of lightweight agents called *ID agents* to various network components for intrusion detection [11]. Each agent is specialized in a certain category of intrusion. For example, the host-based ID agents can analyze BSM audit data, system call traces and user behaviors. On the other hand network ID agents are responsible for network level attacks, such as DDoS and probing attacks. The proposed architecture is hierarchical and it divides the protection and analysis scope into three levels: local, regional and global.

Frincke et al have developed a prototype called the *Hummingbird System* at the University of Idaho [12]. It is a distributed system that employs a set of *hummer* agents, each assigned to a single host or a set of hosts. Each hummer interacts with other hummers in the system through a *manager*, a *subordinate*, and a *peer* relationship. The managers may transmit commands to subordinates. The commands include instructions for gathering or stop gathering data, forward or stop forwarding data etc. Peers may send requests for data forwarding, gathering or receiving to peers. The peer decides whether to honor such requests. The Hummingbird system is intended to allow the administrator to monitor security threats on multiple computers from one central location.

Gopalakrishna and Spafford [1] have presented the architecture of an intrusion detection system that is built with a collection of distributed, autonomous and cooperative agents. In the *interest-based* cooperation and communication model proposed by the authors, the agents request and receive information solely on the basis of their interests. The agents can also specify new interests as a result of a new event or alert. As a major advantage, the entire system is not compromised if an agent fails. Instead, there is a graceful degradation of system performance.

However, most of the hierarchical distributed intrusion detection systems have the following drawbacks [1]:

- *Analysis hierarchy*: As there is a hierarchy in data analysis and data analysis takes place at all levels of the hierarchy, these systems are very difficult to change. In the wake of a new distributed attack, changes may have to be made in modules at many (if not all) levels.
- *Data refinement*: When a module form a lower level sends results of data analysis to a higher level, some data refinement is done. However, the knowledge of what events are important on a system-wide level is circumstantial and is difficult to infer at the lower levels of the hierarchy. If refinement is strict, we may end up losing some system-wide notable events and if the refinement is loose, the higher-level analysis modules will be flooded with large amounts of data from the lower levels. Finding an a priori suitable compromise may be difficult.
- *Bulky modules at all levels of the hierarchy*: Intrusion analysis engines based on anomaly detection are large modules. They consume a significant amount of resources in terms of CPU usage, disk I/O and memory usage, as they have to analyze long audit trails and system state information. In the systems described above, these components are present at all levels of the hierarchy. Such components present multiple points of failure. Degrading or disabling a top-level component would severely limit the detection capability of the system. Moreover, such bulky components are expensive to replicate to achieve fault-tolerance.

- *Passive interactions among modules*: The components of the intrusion detection system interact with each other in a passive way. The lower level components generate data for the upper level components as per the rules driving them. There is no mechanism for a component to query other component on the basis of some analysis that it has done.

The subscription-based communication mechanism between monitors in EMERALD and between the agents and filters in AAFID offers a mechanism for active interaction. But in EMERALD, it is limited to occur between the adjacent levels of the hierarchy and in AAFID, it is allowed only within a host.

Ning *et al.* have recognized the importance of a querying facility in cooperative intrusion detection systems [13][14]. Accordingly, the authors have proposed an extension to the *common intrusion specification language* (CISL) [15] that allows intrusion detection components to specify requests for particular information from other components.

Some authors have proposed the use of mobile agents in intrusion detection. Mell and McLarnon [16] have argued that the major problem with hierarchical intrusion detection systems is due to static locations of the components. They have proposed modeling these components as mobile agents. Helmer and Wong [17] have presented a system that uses lightweight mobile agents for intrusion detection. These agents are dynamically updateable and upgradeable, and due to their smaller size, they are faster to transport. However, mobile agents have serious security concerns and have restricted execution environment. Yang et al have proposed a scheme called CARDS that generates and distributes *detection tasks* among monitors to cooperatively detect attacks [18]. Detection tasks are parts of an attack signature, which the authors refer to as *predefined queries*. However, there is no support for active querying in the proposed architecture.

Thames et al have presented a hybrid intelligent IDS that utilizes a Bayesian network and *self-organizing map* (SOM) [19]. The experimental results have shown that the performance of the hybrid intelligent IDS is better compared to the systems based non-hybrid Bayesian learning approach.

Jemili et al have proposed a framework for an adaptive intrusion detection system that utilizes Bayesian networks [20]. Any new network data that is considered intrusive by the system is added to the dataset of the Bayesian network and the knowledge-base of the system is updated periodically.

Silva et al have implemented an architecture of a remote IDS using the technology of multi-agent systems, web services and *model-driven architecture* (MDA) [21]. The model adapts and extends the concept of network intrusion detection systems, so that the users can use the services provided by the remote IDS without having any IDS on their local hosts. The IDS functionalities are provided as a set of accessible services on the Internet through Web Services.

Ye has presented an agent-based peer-to-peer distributed intrusion detection framework in which functionalities of each agent has been implemented using JACK/UML approach [22]. The knowledge of each agent is represented using an ontological framework. An efficient task allocation protocol is used to coordinate different hosts in the system to collaboratively detect distributed attacks.

Zhao et al have presented a data fusion-based intrusion detection model, in which the detection process is divided into three levels: basic, information and knowledge [23]. The authors have computed an input matrix that introduces Dempster- Shafer theory to the information level of the whole detection so that the results of different detection methods and heterogeneous data in the system can be fused together. Moreover, the intrusion scenario and the system's security situation can be extracted at a higher level.

Zeng and Guo have proposed an agent-based IDS that can be integrated into the applications of enterprise information systems [24]. The system consists of three kinds of agents: the client agents, the server agents, and the communication agents. The agents are integrated to an access control model that enhances the security in the system. To make the system design standardized, the authors have used standard agent design and communication protocols such as *knowledge query and manipulation language* (KQML) [25] and *intrusion detection message exchange format* (IDMEF) [26].

Da Silva et al have proposed an IDS for mobile devices that uses detection mechanisms based on the behavior profiles of the devices and data packets transmitted in a wireless network [27]. This mechanism is particularly suited for resource constrained wireless devices.

For detecting Internet worms, Rasheed et al. have proposed a special-purpose intrusion detection mechanism based on traffic signatures [28]. The system has two algorithms. The first algorithm, called the *intelligent failure connection algorithm* (IFCA) is a worm detection algorithm based on the concept of artificial immune system. The second algorithm, called the *traffic signature algorithm* (TSA) captures the signatures from the network traffic and looks for pattern matching with an Internet worm.

Toutonji and Yoo have presented a novel approach to modeling a worm attack on a computer network [29]. The authors have argued that different parts of a network have different levels of defense requirements and different immunity measures. Accordingly, they have developed a model for splitting the network into two parts: a *highly immune part of the network* (HIN) and a *partially immune part of the network* (PIN). The authors have evaluated the effectiveness of their proposed model by implementing network defense measurements which are adopted from the human immune system. Simulation results have demonstrated that infections due to worms have minimal impact on the HIN.

The proposed approach in this paper is different from the above schemes since it is based on intelligent coordination among the agents and utilizes the reasoning and analytical capabilities of a Bayesian network and an MSBN. The intelligence in cooperation is achieved by communicating events and alerts to only those agents in the system that are interested in those events and alerts. The description of the proposed system is given in the following sections.

## 3. Multiply Sectioned Bayesian Networks

In this section, some background information about Bayesian networks and MSBNs are given. The proposed IDS is based on these concepts and thus a basic knowledge of them is required for proper understanding of the proposed security mechanism.

### 3.1 Bayesian Networks

Bayesian networks are probabilistic models that exploit the conditional independence properties present in a task domain to reduce both the space required to store the model and the time needed to compute posterior probabilities upon receipt of evidence. Formally, we can define a Bayesian network as a graph in which the following conditions hold [30]:

(a) A set of random variables constitutes the set of nodes of the network.
(b) A set of directed edges connects pairs of nodes. The intuitive meaning of an arrow from node $V_i$ to node $V_j$ is that $V_i$ has a direct influence on $V_j$.
(c) Each node has a conditional probability table that quantifies the effects that the parents have on the node. The parents of a node are all those nodes that have arrows pointing to it.
(d) The graph has no directed cycles; hence it is a directed acyclic graph (DAG).

The defining property of a Bayesian network is that the conditional probability of any node given any subset of non-descendants is equal to the conditional probability of that same node given its parents alone. In other words, a Bayesian network represents the exponentially sized *joint probability distribution* (JPD) in a compact manner. Every entry in the JPD can be computed from the information in the Bayesian network by (1):

$$P(x_1,....,x_n) = \prod_{i=1}^{n} P(x_i \mid Parents(x_i)) \quad (1)$$

where, $x_i$'s are the variables and $Parents(x_i)$ represents the parent set of the variable $x_i$ [31].

### 3.2 Multiply Sectioned Bayesian Networks

MSBNs provide a coherent framework for probabilistic reasoning in *cooperative multi agent distributed interpretation systems* (CMADISs) [32]. An MSBN is an extended form of a Bayesian network and consists of a set of interrelated Bayesian subnets that collectively define a Bayesian network. Each subnet encodes an agent's uncertain knowledge about a sub-domain. The subnets in an MSBN are required to satisfy certain conditions so that probabilistic inference can be performed coherently in a modular and distributed fashion [33]. These conditions are as follows:

(a) The subnets in an MSBN must satisfy a *hypertree* condition;
(b) The interface between a pair of adjacent subnets must form a *d-sepset*.

In the following, the above two conditions are illustrated in a more elaborate manner.

**Definition 1.** Let $G = (V, E)$ be a connected graph sectioned into subgraphs $\{G_i = (V_i, E_i)\}$ such that the $G_i$'s can be associated with a tree $\Psi$, with the following property: Each node in $\Psi$ is labeled by a $G_i$ and each link between $G_k$ and $G_m$ is labeled by the *interface* $V_k \cap V_m$ such that for each $i$ and $j$, $V_i \cap V_j$ is contained in each subgraph on the path between $G_i$ and $G_j$ in $\Psi$. Then $\Psi$ is a *hypertree* over $G$. Each $G_i$ is a *hypernode* and each interface is a *hyperlink*.

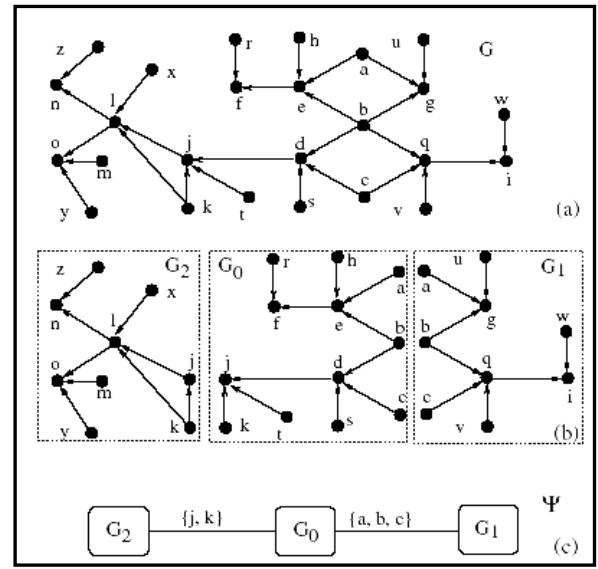

**Figure 1.** The graph $G$ in (a) is sectioned into $G_0$, $G_1$ and $G_2$ in (b). $\Psi$ in (c) is a hypertree over $G$

Figure 1 shows an example. In this example $V_2 \cap V_1 = \phi$ (hence the hypertree condition is trivially satisfied). But in general $V_i \cap V_j$ can be non-empty.

**Definition 2.** Let $G$ be a directed graph such that a hypertree over $G$ exists. Let $x$ be a node that is contained in more than one subgraph and $\Pi(x)$ be its parents in $G$. Then $x$ is a *d-sepnode* if there exists only one subgraph that contains $\Pi(x)$. An interface $I$ is a *d-sepset* if every $x \in I$ is a d-sepnode.

Each of $a$, $b$, $c$, $j$, $k$ in the interfaces of Figure 1 is a d-sepnode. Hence, the interfaces $\{a, b, c\}$ and $\{j, k\}$ are d-sepsets. If the direction of the arc from $j$ to $l$ were reversed, however, the node $j$ would no longer be a d-sepnode and $\{j, k\}$ would no longer be a d-sepset. The hypertree and d-sepset conditions together ensure *syntactically* that the agents can inform each other by passing their beliefs on interfaces only.

In a multi-agent system, a d-sepnode is shared by more than one agent and is called a public node. A node internal to a single-agent is called a private node. Using the concept of d-separation [34], it has been shown that when a pair of subnets is isolated from an MSBN, their d-sepset renders them conditionally independent.

Just as the structure of a Bayesian network is a DAG, the structure of an MSBN is a Multiply Sectioned DAG (MSDAG) with a hypertree organization.

**Definition 3.** A *hypertree MSDAG* $G = \cup_i G_i$, where each $G_i$ is a DAG, is a connected DAG such that (i) there exists a hypertree $\Psi$ over $G$, and (ii) each hyperlink in $\Psi$ is a d-sepset. Figure 1(b) and 1(c) show a hypertree MSDAG.

**Definition 4.** An MSBN $M$ is a triplet $(V, G, P)$. $V = \cup_i V_i$ is the *domain* where each $V_i$ is a set of variables, called a *sub-domain*. $G = \cup_i G_i$ (a hypertree MSDAG) is the *structure* where nodes of each DAG $G_i$ are labeled by elements of $V_i$. Let $x$ be a variable and $\Pi(x)$ be the parents of $x$ in $G$. For each $x$, exactly one of its occurrence (in a $G_i$ containing $\{x\} \cup \Pi(x)$) is assigned P (x | $\Pi(x)$), and each occurrence in other

DAGs is assigned a uniform potential. $P = \Pi_i\ P_i$ is the *joint probability distribution* (JPD), where each $P_i$ is the product of the potentials associated with the nodes in $G_i$. A triplet $S_i = (V_i, G_i, P_i)$ is called a *subnet* of $M$. Two subnets $S_i$ and $S_j$ are said to be adjacent if $G_i$ and $G_j$ are adjacent.

MSBNs form a coherent framework for probabilistic reasoning in CMADISs. Each agent holds its partial perspective of a large problem domain, accesses a local evidence source, communicates with other agents infrequently, reasons with the local evidence and limited global evidence, and answers queries or takes actions. It has been shown that if all agents are cooperative and each pair of adjacent agents (belonging to adjacent subnets) are conditionally independent given their shared variables and have common initial belief on the shared variables, then a joint system belief is well-defined, which is identical to each agent's belief within its sub-domain, and supplemental to the agent's belief outside the sub-domain. Even though multiple agents may acquire evidence asynchronously in parallel, the communication operations of MSBNs ensure that the answers to queries from each agent are consistent with the evidence acquired in the entire system after each communication. Since communication is infrequent, the operations also ensure that between two successive communications the answers to queries for each agent are consistent with all local evidence gathered so far, and are consistent with all evidence gathered in the entire system up to the last communication. Therefore, an MSBN can be characterized as one of functionally accurate, cooperative distributed system [35]. In Section 5, it will be seen how an MSBN can be applied to make distributed probabilistic inferences

## 4. Architecture of the System and the Agents

In this section, the architecture of the overall system is described. In particular, the architecture of the agents is described in detail. The agents collaborate with each other to make an efficient distributed intrusion detection framework.

### 4.1 System Architecture

The proposed security mechanism is a distributed, lightweight, agent-based intrusion detection system. The model architecture is similar to what has been proposed in [36], but differs completely in the mechanism of *trust management* and *fault-tolerance*. In the proposed approach, the agents are viewed as autonomous, reflexive, proactive and cooperative entities. They are responsible for collecting data, analyzing them, and making appropriate inference form the analysis. The agents use an inference process that utilizes the collected data as evidences in a Bayesian network. Monitoring and analysis work is duplicated for accuracy and fault tolerance, e.g., handling the possible situations when some agents are compromised.

The agents are grouped into several sub-domains. The agents in the same sub-domain communicate actively and frequently. Communication between agents belonging to adjacent sub-domains happens infrequently. The agents have knowledge about a Bayesian network model of the structures of well-known attack types as well as normal usage pattern, which is constructed offline from data repositories containing system logs from ongoing attacks. This global Bayesian network has been partitioned into multiple subnets based on the spatial locations of the agents. The agents in the same sub-domain have the common knowledge of the subnet in their sub-domain. Each agent is delegated with responsibilities to monitor certain predefined security parameters (attack signatures) at the sub-domain in which it resides. In addition, some agents are responsible for monitoring the network traffic data to detect possible network-based attacks.

Bayesian networks are used to represent the existing knowledge of different attack signatures [37]. A Bayesian network is capable of capturing the mutual influence of different domain variables on target attributes. Using a Bayesian network model one can infer the probabilities of occurrence of different intrusion types, which are easy for human security investigators to interpret. Moreover, this representation can easily accommodate prior domain knowledge. Bayesian network approach also allows for combining two different intrusion detection methodologies such as anomaly detection and signature recognition [38]. To facilitate this process in the proposed mechanism, one Bayesian network is generated whose nodes classify several known attack types and normal system behavior. Using this network, the agents can detect both normal behavior and a known attack type. If the probabilities associated with none of the target nodes cross the threshold, given the input feature values, an anomalous behavior is suspected.

As described in Section 6, every host in the system has one special agent, called the *distributed trust manager* (DTM), which continuously sends messages to its peers in other hosts. By applying the *Byzantine agreement protocol* (BAP) among the peer hosts, the system can identify a host that is possibly compromised by a malicious intruder, and isolate that host from intrusion detection process. This distributed trust mechanism makes the proposed system robust and fault-tolerant.

### 4.2 Agent Architecture

Figure 2 depicts the architecture of an agent. Each agent consists of six modules. The functionalities of each of these modules are briefly described below.

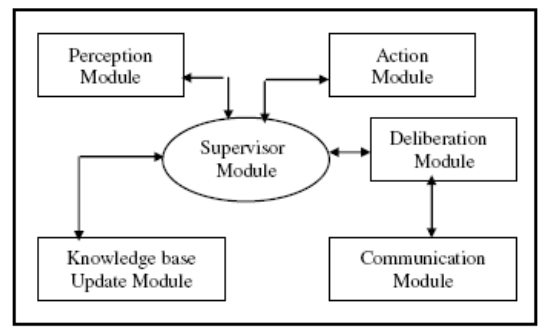

**Figure 2.** The architecture of an agent embedded in a node

- **Perception module:** This module is responsible for collection of audit or network data of the sub-domain (or subnet) to which the agent belongs.
- **Deliberation module:** This module is responsible for analyzing the data collected by the Perception module. Essentially, its role is to enable the agents to reason and extrapolate by relying on built-in knowledge (beliefs) and experience in a rational way. Decisions of the agents depend on the security environment status, and collected evidence. It also allows an agent to update its belief associated with the node of the subnet it is monitoring.
- **Communication module:** This module allows an agent to communicate its belief, decisions, and knowledge to its peer agents. The inference made by an agent is passed to its peers in the same sub-domain, and possibly, to other sub-domains also.
- **Action module:** The role of this module is to take appropriate actions when a possible intrusion is detected. When an agent recognizes that a monitored host is exceeding the threshold of one known attack, it triggers an alert for that particular attack and communicates it to the system administrator. Besides this signature-based detection, the agents can also trigger an alert indicating 'an anomalous situation', when the activated target node does not belong to those representing normal behavior or any of the known attack types considered while the Bayesian network was constructed. The system administrator can either confirm the attack (or take necessary steps to handle it), or reject the alert if it is found to be a false alarm on further probing.
- **Knowledgebase update module:** If the system administrator confirms an anomaly alert, a Bayesian network is modified to accommodate this new attack in the knowledge base. This attack is now recognized as a known attack and will be considered as an attack signature for future monitoring purpose. Thus the agents are adaptive to the discovery of new types of network intrusions.
- **Supervisory module:** This is the central module that coordinates the tasks and interactions among the other modules described above.

## 5. Communication and Security

Three types of agents are deployed in the proposed system. Figure 3 shows the interactions among different types of agents.

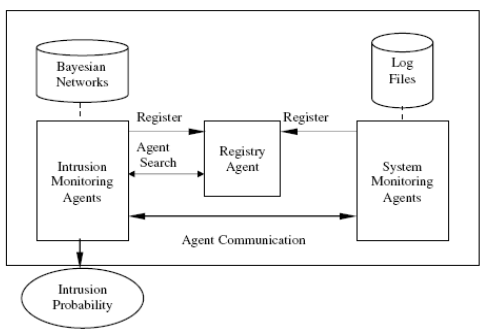

**Figure 3.** The major components of the system

The *system monitoring agents* are responsible for collecting, transforming, and distributing intrusion specific data upon request and evoke information collection procedures. These agents publish the details of the variables they monitor, which can be utilized by other agents. The *intrusion monitoring agents* subscribe to beliefs published by the system monitoring agents and other intrusion monitoring agents. Each intrusion monitoring agent has a local knowledge about a Bayesian network structure of attack types. These agents update their beliefs on receipt of information from other agents. For each registered agent, a registry maintains information about the monitored variables. The agents use the registry to find information (e.g., name and location) about agents that may supply required data. Once location and the name of an agent providing required data are found, the registry need not be referred again. The messages exchanged between the agents are in *extensible markup language* (XML). The important messages exchanged are: i) registration of agents with registry agents, ii) request to registry agents for finding the locations of other agents, iii) search of agent queries, iv) belief subscription requests, v) belief update messages.

There are two categories of communications among agents: communications among agents residing at the same host, and communication among agents on different hosts. Figure 4 shows these two different types of communications. Different mechanisms for these two types of message communications have been proposed and compared in some works [38][39][40]. In case of communication among agents in the same host, the agents communicate using methods like pipes, message queues, and shared memory. In the proposed mechanism, a shared memory architecture is used for agent communication since it allows large volume of data to be shared among agents [8]. In case of communication among agents over the network, it is not worthwhile to replicate the same method of communication for each agent. Instead, an *agent management system* (AMS) which enables communication between different agents for intra-host and inter-host communication will be a much more efficient scheme. In the prototype system developed, the capabilities provided by the *java agent development environment* (JADE) [39] environment have been utilized to build an AMS.

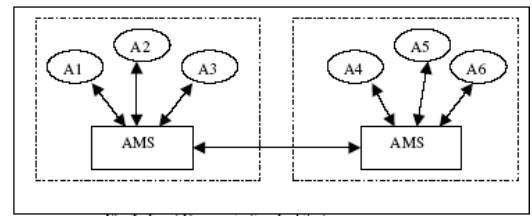

**Figure 4.** The architecture of agent communication

All the message communications among the agents are made secure by incorporating cryptographic mechanisms. *Public Key Infrastructure* (PKI) is used to provide two-way authentication of agents and messages. The messages communicated among the agents are all encrypted using private key encryption mechanism.

## 6. Intrusion Detection using Bayesian Hypothesis

This section describes how the concepts of Bayesian networks and MSBNs are applied in the domain of intrusion detection with the help of a set of cooperative agents. The agents are grouped so that they form a Bayesian network and a distributed inference mechanism is developed among them with the help of an MSBN.

### 6.1 Bayesian Network in Intrusion Detection

In Section 3, a Bayesian network has been defined as a directed acyclic graph (DAG) with the nodes representing the variables, and each directed edges representing a dependency between the corresponding variables. The effect of the parents of a node on a node is represented by conditional probabilities of that variable given values of its parent nodes in the form of a *conditional probability table* (CPT). In security domain, it is useful to represent a set of attack signatures by a Bayesian network for the following reasons. Firstly, a Bayesian network can handle incomplete information. In most of the case, the agents may have limited local view of the network and may receive only partial information about a possible attack. Secondly, a Bayesian network can represent causal relationships among variables, which can help an intrusion detection model to combine a priori knowledge and observed data to take a decision. Lastly, a Bayesian network allows updating of the beliefs and thus can be used to recognize novel attack signatures by the intrusion detection system. In the proposed security system, a Bayesian network is first constructed from a database of known attacks. This network is then partitioned into several sub-trees following the principle of MSBNs [4], and distributed among the agents. The agents belonging to different subnets communicate among themselves, update their beliefs about different events, and carry out a distributed intrusion detection activity [41].

### 6.2 Inference with MSBNs

An MSBN consists of a set of interrelated Bayesian subnets each of which encodes an agent's knowledge of a sub-domain. In such a framework, probabilistic inference can be performed in a distributed fashion, while answers to queries are exact with respect to probability theory. Existing methods for multi-agent inference in MSBNs are extensions of a class of methods for inference in single-agent Bayesian networks: message passing in junction trees [3][37][42]. The *linked junction forest* (LJF) method [43][44] compiles each subnet of a multiply connected network into a *junction tree* (JT), by clustering the triangulated moral graph of the underlying undirected graph. The algorithm performs message propagation over the JT. Message passing among agents belonging to two adjacent subnets is performed through a *linkage tree* between a pair of adjacent nodes each belonging to adjacent subnets. Though the exact belief update in a Bayesian network is NP-hard [45][46], it can still be used for the purpose of intrusion detection since the subnet sizes are usually small.

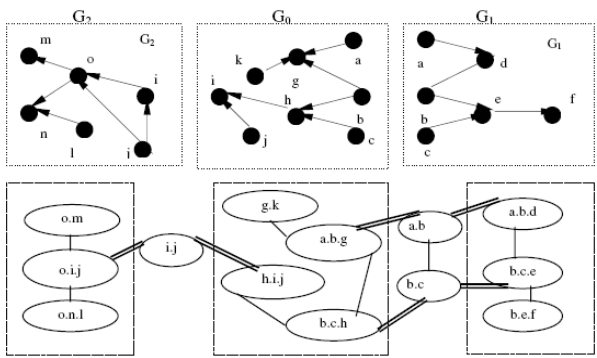

**Figure 5.** The DAGs of the three subnets of an MSBN and JTs constructed from the subnet

Figure 5 shows an MSBN with three subnets $G_0$, $G_1$, $G_2$. Each of these subnets contains a group of agents. The knowledge of a group of agents is encoded in the corresponding subnet to which it belongs. The LJF method has compiled each of three subnets into a JT (called a local JT), and has converted each d-sepset into a JT (also called linkage tree). Figure 5 also depicts three local JTs and two linkage trees of the monitoring system. Each oval in a JT or a linkage tree represents a subset of variables and is called a *cluster*. For instance, {$o, i, j$} is a cluster in a JT in the subnet $G_2$ and {$i, j$} is a cluster in the linkage tree between subnets $G_2$ and $G_0$. Local inference is performed by message passing in the local JT. Message passing between a pair of adjacent sub-domains is performed using the linkage tree.

Once a multi-agent MSBN is constructed, agents may perform probabilistic inference by computing the query $P(x|e)$, where $x$ is any variable within the sub-domain of a group of agents, and $e$ is the observations made by all the agents in the system. The key computation is to propagate the impact of observations to all the agents in the entire system. This system-wide communication among the agents is vital for sharing of information among agents belonging to different sub-domains. As the agents are designed to be autonomous, the need for system-wide message passing arises infrequently. Most of the time, the agents in subnet $G_i$ computes the query $P(x|e_i, e_i')$, where $e_i$ is the local observations made by the agents in $G_i$, and $e_i'$ is the observations made by the agents of other sub-domains as recorded in $G_i$ till the last communication. This computation is called *local inference*. It has been proved that among different distributed multi-agent inference algorithms in MSBNs, LJF has least overhead of inter-agent communication [33]. Thus use of LJF method ensures that the network traffic due to the security mechanism is kept to a minimum.

## 7. Fault- Tolerance and Trust Mechanism

In this section, a novel approach for introducing fault-tolerance in the proposed system is described. A distributed trust management scheme is developed among the agents in the system and a robust algorithm based on *Byzantine agreement protocol* (BAP) [47] is invoked among the peer agents. This enables a reliable and fast detection of any compromised agents in the system. If any agent is detected to be compromised, it is immediately isolated from the system. This makes the intrusion detection mechanism reliable, secure and fault-tolerant.

### 7.1 Distributed Trust Management

The agents in a distributed intrusion detection system are always vulnerable to attacks by intruders. If an intruder can compromise any host in the system, the detection capability of the entire system will be severely affected. The agents in a compromised host will attempt to influence the JT and their effect will be propagated in the entire system by the message passing mechanism among the agents unless the compromised host is detected and isolated promptly. To ensure early detection of any compromised host(s) in the system, an efficient trust management scheme based on *Byzantine Agreement Protocol* is developed among the peer hosts.

### 7.2 The Byzantine Generals Problem

Lamport et al described the *Byzantine generals problem* in [47]. Essentially, the problem formulation is as follows: Imagine that several divisions of a Byzantine army are camped outside an enemy city, each division commanded by its own general. The generals can communicate with each other only by messengers. After observing the enemy, they must decide upon a common plan of action. However, some of the generals may be traitors, trying to prevent the loyal generals from reaching agreement. The generals must have an algorithm to guarantee that all loyal generals decide upon the same plan of action. The loyal generals will all do what the algorithm asks them to do, but the traitors may do anything they wish. The loyal generals should not only arrive at an agreement but should agree upon a reasonable plan.

Lamport devised two solutions to this problem –the *Oral Message Algorithm* (OMA) and the *Signed Message Algorithm* (SMA) [47]. The OMA requires more than $2^n$ messages to be sent for achieving the consensus, if there are *n* generals and it works only if the number of loyal generals is greater than twice the number of traitors [47]. The SMA, in contrast, requires only $O(n^2)$ messages to achieve consensus, and works effectively if there are at most *n - 2* number of traitors [47]. Moreover, n comparison to OMA, SMA works faster. However, as discussed in Section 6.3, SMA requires more number of necessary conditions to be met by the system.

### 7.3 Byzantine Agreement Protocol – Signed Message Algorithm

The Byzantine agreement protocol (BAP) is essentially an algorithm designed to achieve consensus among nodes in a distributed system. A set of processes can arrive at a consensus if they all agree on some allowed values called the 'outcome' (if they could agree on any value the solution would be trivial: always agree on 0). Thus arriving at a consensus involves two actions: first specify a value, and the read the outcome of execution of the processes involved. The consensus algorithm terminates when all non-faulty (not compromised) processes come to know the outcome. If we consider the generals in BAP as the hosts in a distributed system, and the consensus as the requirement of agreement among the hosts as which agents are safe/sane (i.e. not compromised), then the problem of identifying and isolating any compromised host(s) in a distributed can be described more formally as follows.

Consider a distributed system consisting of several hosts with each host having a set of agents running on it. The agents cooperate to detect intrusions into the system. Each host runs a special agent, called the *distributed trust manager* (DTM), which continuously sends messages to its peers on other hosts. The message can be of two types: i) *Message $A_1$:* The host is safe (i.e. not compromised), with a value "0". ii) *Message $A_2$:* "The host is compromised with a value, "1". The signature of the possible intrusion also may be sent along with this message.

The Signed Message Algorithm will work correctly if we can guarantee the following:

- Every message sent by any host is delivered correctly.
- The receiver of a message knows who the sender is.
- The absence of a message in the buffer of a host can be detected.
- The signature of a legitimate (i.e. not compromised) host cannot be forged; any alteration made on a signed message can be detected.

Any host can verify the authentication of the signature of its general.

With cryptographic mechanisms based on Public Key Infrastructure (PKI), it is possible to ensure all the conditions stated above, and thus SMA can be assumed to work well in a security framework based on cryptography.

**Algorithm:** *Signed_Message(m)*

Initially $V_i = \phi$

[1]  The commander signs and sends its value to every host $H_i$ that it can reach directly.

[2]  For each *i*
  (A) if the host $H_i$ receives a message of the form *v* : 0 from the commander and it has not received any order, then
    (i) it lets $V_i$ equal {*v*};
    (ii) it sends the message *v* : 0 : *i* to every other host.
  (B) if the host $H_i$ receives a message of the form *v* : 0 : $j_1, j_2,\ldots j_k$ and *v* is not in the set $V_i$, then
    (i) it adds *v* to $V_i$,
    (ii) if $k < m$, then it sends the
message *v* : 0 : $j_1, j_2,\ldots j_k$: *i* to every host
    other than $H_{j1}, H_{j2}, ..H_{jk}$.

[3]  For each i: When the host $H_i$ will receive no more messages, it obeys the order *choice*($V_i$).

In the Signed Message Algorithm, one of the hosts acts as the leader and sends an order to the other hosts. Whenever a host receives a message, it takes the order and puts it in its list of the orders received. Then the receiver signs the message with its own signature and forwards it to all the hosts whose signature is not on the order. If a host receives a message with an order that is already in his list, he ignores the message. When no more messages are left to be received, all the hosts choose an order from the list of orders they have received using this method. If only one order has been received, then that order is chosen. Because any order that reaches a loyal general will be forwarded to all other generals who have not seen the order, all the loyal generals will have the same set of orders to choose from, and thus choose the same order to obey.

### 7.4 Distributed Trust Manager

The role of the DTM is described in this section. The concept of DTM has been borrowed from [48], where it is utilized for establishing and managing trust in a distributed environment. DTM is largely responsible for forming and maintaining *trust domains*. A trust domain is a set of hosts that share a charter and a security policy, and behave consistently in accordance with that policy. The hosts in a trust domain work in collaboration to prevent compromised hosts from joining the trust domain. If any host becomes compromised after joining a trust domain, other hosts in the domain will be able to detect it and isolate it.

DTM uses a consensus algorithm among the members of the trust domain to perform its task. We assume that at the beginning of a trust domain formation, all the hosts in the trust domain are sane. In other words, initially, none of the hosts in a trust domain are compromised. DTM tries to detect and remove any host that becomes compromised after it has joined the trust domain. Any compromised host in the trust domain is identified by running $n$ instances ($n$ is the max number of hosts in the trust domain) of the *Signed_ Message* algorithm in parallel, assuming that the majority of the hosts in the trust domain are not compromised. If the *leader* of the *Signed_Message* algorithm is not compromised, then after running the algorithm in parallel, all the host that are not compromised will know that the leader is not compromised. If the leader of an execution of the *Signed_Message* algorithm is compromised, then any of the following situations will occur:

- The leader sends *0* messages to all hosts that are not compromised. In this case, all the hosts that are not compromised will assume the host (leader) to be compromised or dead.
- The leader sends *1* message to only some of the hosts that are not compromised. In this case, the hosts that are not compromised and receive *1* message from the leader are able to detect that there is a compromised host in the system. These hosts, then, send messages to other hosts in the system accusing the suspected compromised hosts. The accused hosts are then tested, and are determined to be compromised or not compromised.
- The leader sends *1* message to all the hosts that are not compromised. All the hosts that are not compromised understand that the message is wrong, and the host (leader) is compromised, if it contradicts the majority. If the message does not contradict the majority, the leader cannot be detected to be compromised, unless it sends a different message to at least one compromised host, which in turn forwards the message to another host that is not compromised. Although the leader should be removed in this case, it is not a critical problem, as it is not causing any damage to the system at present.
- The leader sends two (or more) different messages to some hosts that are not compromised. All the hosts that are not compromised see contradictory instructions, and understand that the host (leader) is compromised.

It is, therefore, observed that DTM can identify compromised hosts in the system in all possible cases, and isolate it from the trust domain.

## 8. Experiments and Results

A proof-of-concept prototype for the proposed IDS has been built using Java and JADE [39]. JADE is a middleware developed by Telecom Italia Lab (TILAB) for enabling faster development of multi-agent distributed applications based on the peer-to-peer communication architecture. JADE has been implemented fully in Java. It includes both the libraries (i.e. the Java classes) required to develop application agents, and the run-time environment that provides the basic services and that must be active on the host before agents can be executed. From the functional point of view, JADE provides all the basic services necessary for distributed peer-to-peer applications. It allows each agent to dynamically discover other agents and to communicate with them by message passing mechanism. The agents communicate by exchanging asynchronous messages- the communication model universally accepted for distributed systems. Each agent is identified by a unique identifier and provides a set of services. An agent can register its services and search for other agents providing given services. It can control its life cycle also.

In the prototype developed, each agent is endowed with three behavioral capabilities: *filtering*, *interaction*, and *deliberation*. The filtering behavior of an agent enables it to filter security events from the observations it makes. When an event occurs in the network, it is collected by an agent only if it matches with the event classes specified in the detection goal of the agent. The interaction behavior manages the interaction between different agents. It defines the mailbox of the agent, and the way the messages are received and enqueued for later interpretation. The deliberation behavior of an agent allows it to represent its beliefs, goals, intentions, and knowledge in a semantic format. When an agent receives a detection goal, it updates a set of event classes to filter. When an event occurs, it is filtered by the filtering module and sent to the deliberation module. The deliberation module updates /creates the agent's beliefs, and tests whether the belief matches with an attack signature. If it matches, then a detection goal is reached and a list of intentions is sent to the interaction module for execution.

Essentially, development of agents under JADE framework involves the following steps of activities:
- Determination of the agent behaviors.
- Implementation of the agent class (extending the existing classes of JADE).
- Implementation of the agent meta-behavior by instantiating an existing class or introducing a new class and then instantiating it. The meta-behavior provides an agent with a self-control mechanism to dynamically schedule its behaviors in accordance with its internal state.
- Instantiation of the agent class.
- Initializing the agent acquaintances.
- Deployment and activation of the agent.

Some experiments have been conducted to test the performance of the developed prototype. In the experiments, the KDD Cup 1999 intrusion detection contest data [49] has been used. This data was compiled during 1998 DARPA intrusion detection and evaluation program by MIT Lincoln Lab [50]. The original data contains 744 MB of information with 4.94 million records. The dataset has 41 attributes for each connection record plus one connection record specifying one of 24 different types of attacks or normal condition. Thus, effectively each record is given a class label that specifies the category of attack to which the record belongs. All the attacks are grouped into 4 major categories: (i) *denial of service* (DoS), (ii) *remote to local* (R2L), (iii) *user to root* (U2L), and (iv) *probe*.

We constructed a dataset consisting of 15000 records by randomly selecting records from the original database, such that the number of data instances selected from each class was proportional to their frequencies in the original database. We added one more class of records that we call 'normal' class apart from the 4 attack types mentioned above. The knowledge about these attacks is then distributed among the agents in the system. A Bayesian Network Power Constructor (BNPC) [51] is used to generate a Bayesian network from these sampled records. This sample will contain both attack signatures and normal records (i.e. the event sequence that does not constitute an attack). This Bayesian network is sectioned into multiple subnets utilizing the rules for sound partitioning in MSBN [4]. Finally, the LJF method [4] is used for intrusion detection.

**Table 1.** Impact of the IDs on User Applications

| Number of users | 10 | 20 | 30 | 40 | 50 |
|---|---|---|---|---|---|
| Memory required with IDS (K) | 331 | 471 | 601 | 710 | 795 |
| Memory. required without IDS (K) | 324 | 453 | 572 | 669 | 740 |
| CPU usage with IDS (%) | 53 | 67 | 80 | 86 | 92 |
| CPU usage without IDS (%) | 48 | 63 | 75 | 82 | 87 |

The performance of the prototype has been tested in a network of 50 workstations, with each workstation having Pentium 4 processor, 3GHz clock speed, 1 GB RAM, and Red Hat Linux version 9 as the operating system. The data-rate of the Ethernet was 100 MBPS. The attack knowledge base was distributed among the agents in the workstations in the form of a MSBN as described earlier.

The evaluate the performance overhead of the IDS, the average memory space required and the average CPU usage on the workstations were observed with varying number of active users. For this purpose, the number active users were varied from 10 to 50 and the average load on the workstations was noted. Table 1 summarizes the results. It is evident that the average memory and the CPU usage on a workstation due to the IDS were marginal and the overhead decreased with the increase in the number of users. This makes the system scalable.

To test the CPU utilization by the agents, some attacks are simulated on the workstations and in the network. The attacks included guess_passwd, buffer_overflow, portsweep, teardrop, mailbomb, etc.

During the thirty-minute analysis period, the maximum CPU utilization of the agents was found to be only 8.76%, the average utilization of the entire period being 5.34%. It is evident that the proposed IDS has a fairly low memory and computational overhead.

Using *Ethereal* network *sniffer* [52] (a software to capture and analyze information being transmitted over a network), the network was monitored and the bandwidth consumption of the agents was evaluated. First, the network traffic volume was observed without the IDS agents running on the workstations and then the agents are activated on all the 50 workstations and traffic volume in the network was monitored again. From the data collected by the sniffer, it was evident that the agents had very little bandwidth consumption.

During the one-hour time when the agents were active on the workstations, the sniffer found only 15% increase in the number of packets in the network as compared with the one-hour period without the agents. The average bandwidth consumption by the agents never exceeded 5% of the 100 MBPS Ethernet.

**Table 2.** Operational Performance of the IDS Prototype

| Activity Type | Detection Rate (%) | | False Positive (%) |
|---|---|---|---|
| | Proposed IDS | Li et al IDS | Proposed IDS |
| DoS | 98.25 | 97.57 | 10.25 |
| R2L | 7.31 | 0.37 | 12.43 |
| U2R | 86.42 | 71.49 | 10.57 |
| Probe | 94.28 | 90.49 | 11.87 |
| Normal | 97.80 | 98.13 | 7.31 |

For testing the detection efficiency and the false positive rates of the proposed IDS, 37 different attacks are simulated on the workstations and on the network. The victim and the attacking workstations are chosen randomly so as to test all the part of the network. While the attacks are simulated in the network and the workstations, some of attacks are chosen in such a way that they are not in the knowledge base of the agents in the IDS. This is done to test the ability of the IDS to detect novel attacks.

The detection efficiency of the proposed IDS is also compared with a recently proposed scheme by Li et al. [53]. From Table 2, it is evident that except for the 'normal' category, the detection efficiency of the proposed IDS is better than the scheme proposed by Li et al. [53].

The better performance of the proposed scheme is due to the robust knowledge base building and inference mechanism of the JADE environment. However, like most of the existing IDS schemes, the proposed scheme has a low detection rate for R2L attacks. Although, the detection rate for these attacks is much higher compared to Li et al.'s scheme, it is far from satisfactory. Since R2L attacks are essentially different from the other types of attacks, there is a need for a different approach to the detection logic design for these types of attacks. The DTM scheme is mostly responsible for higher detection rates in DoS and probe attacks. The false positive rates for all attack types were found to be fairly low as may be observed from Table 2.

Finally, the performance of the trust management system is studied. For this purpose, a group of nodes is chosen randomly for formation of the trust domain.

Once the formation of the trust domain is done, attacks are simulated on some of the nodes in the trust domain from outside the network. The DTM in each node is activated and time required to identify and isolate the compromised nodes are studied. This experiment is conducted 50 times, and in all occasions it has been observed that the compromised nodes in the trust domain are identified and isolated from the network activities by the other nodes in the trust domain. The maximum time for detection and isolation was observed to be 35 s and the average time being 15 s. The maximum time was observed when the network was having a very heavy traffic and all the nodes were busy with intensive applications running on them. The results thus show that the trust management system is also very effective and efficient.

## 9. Conclusion

In this paper, the framework of a distributed IDS is presented. The IDS consists of a set of autonomous agents that cooperate with each other to carry out a distributed intrusion detection process. Using distributed computation and message passing between the agents, the IDS can detect both signature-based attacks and anomalous activities in real-time. Apart from its ability to make distributed inference based on multiply sectioned Bayesian networks, the proposed IDS can also identify and isolate any compromised nodes in the system with the help of Byzantine Agreement Protocol among the peer nodes. The experiments conducted on the prototype of the system have shown that the detection efficiency and the false positive rates of the proposed IDS is better than some of the currently existing scheme. Development of a new detection logic for R2L types of attacks constitutes a future plan of work.